\documentclass{jfm}
\usepackage{amsmath, mathtools, mathrsfs, amssymb, mathabx}
\usepackage{xfrac}
\usepackage{bm}
\usepackage{graphicx}
\usepackage{epstopdf, epsfig}
\usepackage[dvipsnames]{xcolor}

\shorttitle{Generation and decay of yawed wind turbine counter-rotating vortices}
\shortauthor{C. R. Shapiro, D. F. Gayme and C. Meneveau}

\title{Generation and decay of counter-rotating vortices downstream of yawed wind turbines in the atmospheric boundary layer}

\author{Carl R. Shapiro\aff{1}
  \corresp{\email{cshapir5@jhu.edu}},
Dennice F. Gayme\aff{1}
 \and Charles Meneveau\aff{1}}

\affiliation{\aff{1}Department of Mechanical Engineering, Johns Hopkins University, Baltimore, MD 21218, USA}

\begin{document}

\maketitle

\begin{abstract}
A quantitative understanding of the dominant mechanisms that govern the generation and decay of the counter-rotating vortex pair (CVP) produced by yawed wind turbines is needed to fully realize the potential of yawing for wind farm power maximization and regulation. Observations from large eddy simulations (LES) of yawed wind turbines in the turbulent atmospheric boundary layer and concepts from the airplane trailing vortex literature inform a model for the shed vorticity and circulation. The model is formed through analytical integration of simplified forms of the vorticity transport equation. Based on an eddy viscosity approach, it uses the boundary layer friction velocity as the velocity scale  and the width of the vorticity distribution itself as the length scale. As with the widely used Jensen model for wake deficit evolution in wind farms, our analytical expressions do not require costly numerical integration of differential equations. The predicted downstream decay of  maximum vorticity and total circulation agree well with LES results. We also show that the vorticity length scale grows linearly with downstream distance and find several power laws for the decay of maximum vorticity. These results support the notion that the decay of the CVP is dominated by gradual cancellation of the vorticity at the line of symmetry of the wake through cross-diffusion. 
\end{abstract}

\begin{keywords}
\end{keywords}

\section{Introduction}
The spanwise component of a yawed wind turbine's axial force induces a counter-rotating vortex pair (CVP) that laterally deflects and deforms~\citep{Branlard2016a, Bastankhah2016a, Howland2016a} its wake downstream. This phenomenon has the potential to increase or regulate wind farm power output~\citep{Howland2019a}. Fully harnessing this potential requires a rigorous understanding of the underlying fluid dynamics. Efficient engineering prediction methods of the mechanisms governing the generation and decay of the induced vorticity downstream of the yawed turbine in the atmospheric boundary layer (ABL) enable wind farm design and operational decisions that take advantage of this knowledge.

The fate of strong streamwise vortices in the ABL, such as the yawed wind turbine CVP, has been studied extensively. Aircraft wings at takeoff generate counter rotating tip vortices that can stay near the runway and generate dangerous conditions for the next takeoff~\citep{Spalart1998a,Gerz2002a}.
From a fundamental fluid dynamics viewpoint, much effort has been invested in understanding the decay process of vortices in turbulent flow~\citep{Tombach1973a,Devenport1996a,Jaarsveld2011a,Takahashi2005a}. In the case of yawed wind turbines, the vast literature on aircraft trailing wake vortices and the individual helicoidal vortices shed by individual turbine blades~\citep{Ivanell2010a,Sorensen2011a,Chamorro2013a} is useful as a conceptual guide. However, this literature is  not directly relevant to the large-scale CVP shed by yawed wind turbines. Their CVP vortex core is expected to scale with the turbine diameter, rather than the chord length of each blade, and their circulation is significantly weaker than that of aircraft trailing vortices since the overall sideways forces generated by the blades sweeping the inclined turbine disk area is only a fraction of the total turbine axial force. 
 
Recent work is just beginning to link the yawed wind turbine CVP to the airplane trailing vortex literature: Treating the yawed wind turbine as a porous lifting surface and applying Prandtl's lifting line theory, our recent theory predicts the initial magnitudes of the transverse velocity and the circulation of the shed CVP~\citep{Shapiro2018b}. From this insight, recent work has treated the initial vorticity distribution as point vortices along the edge of the swept area of the rotor~\citep{Martinez2019b, Zong2020a, Martinez2020a} that diffuse under turbulent mixing, i.e. Lamb-Osseen vortices~\citep{Saffman1992a}. The diffusion rate is specified by an eddy viscosity that is determined empirically~\citep{Zong2020a} or using a mixing length model with the velocity scale specified by the wake velocity gradient and mixing length specified by the size of the largest ABL eddies~\citep{Martinez2019b}. The downstream evolution is then found by numerically integrating the resulting vortex system. This numerical approach yields results that agree well with simulations and experiments, but does not facilitate insight into fundamental vorticity decay mechanisms or reveal scaling laws based on the turbine yaw angle or the ambient turbulence characteristics. 

In this work, we study the generation and decay of the CVP generated from yawed wind turbines in the ABL. In order to advance engineering models for the shed vorticity, analogous to the Jensen model~\citep{Jensen1983a} for the velocity deficit, we seek to derive analytical expressions that do not require numerical integration. Our model is motivated and validated by large eddy simulation (LES) data, discussed in~\S\ref{sec:les}, and the trailing vortex literature. In~\S\ref{sec:generation}, we analytically derive the vorticity, transverse velocity, and circulation distribution generated immediately downstream of a yawed actuator disk and compare the analytical predictions to simulations. In~\S\ref{sec:decay}, an eddy-viscosity assumption is applied to model the turbulent diffusion during the downstream evolution of this initial vorticity distribution.  We propose appropriate velocity and length scales to be used to define an eddy-viscosity that reproduces LES measurements. We derive analytical expressions for the maximum vorticity and total circulation of each vortex and compare these to LES. Of particular interest is to establish whether the decay of the CVP vortex strength can be explained by a simple model of cross-diffusion between the two vortices. 

\section{Large eddy simulations of yawed wind turbines in the ABL.}
\label{sec:les}

\begin{figure}
\centerline{\includegraphics[width=\textwidth]{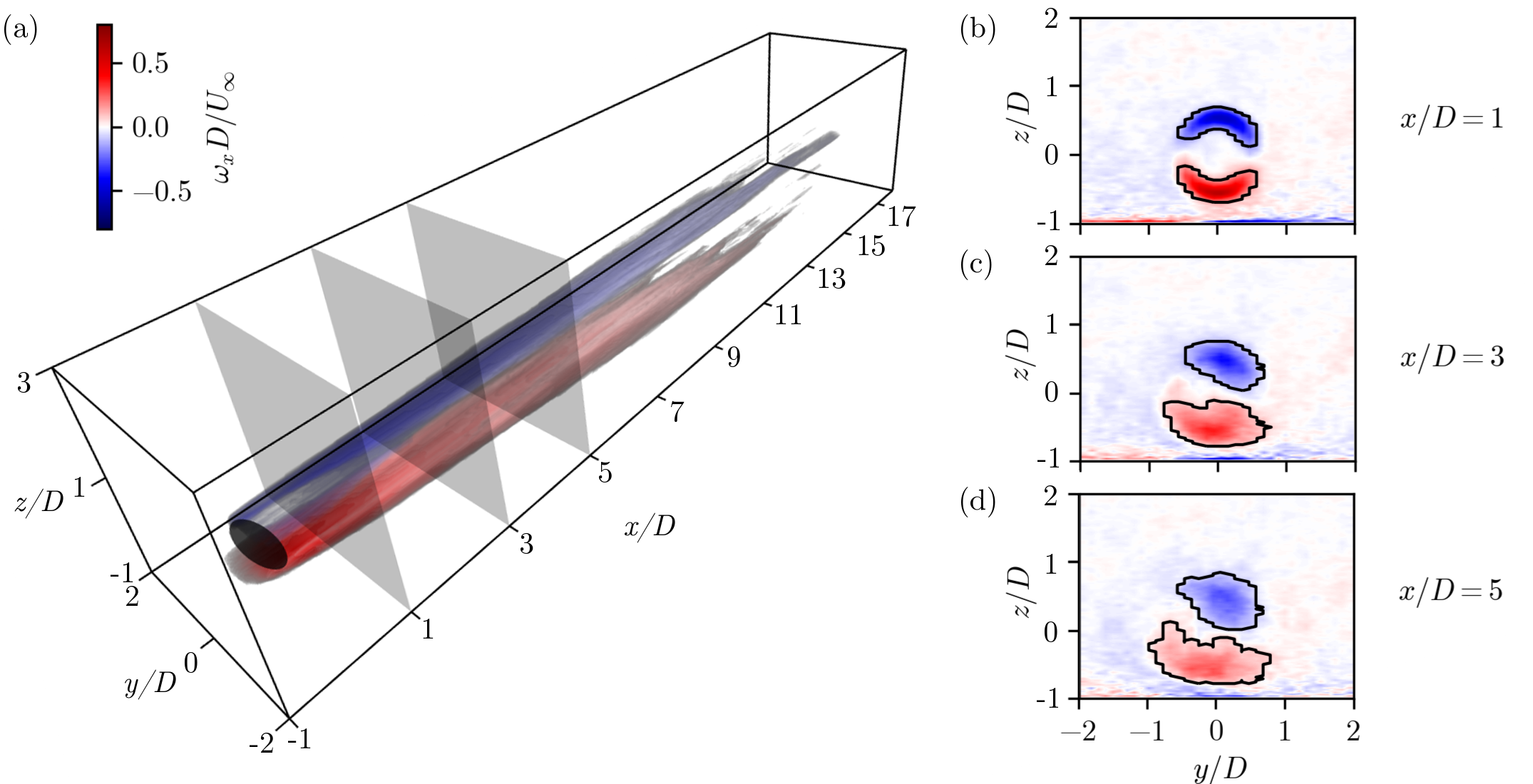}}
\caption{Time-averaged streamwise vorticity distribution behind a yawed wind turbine $\gamma = 20^\circ$ under turbulent ABL inflow. (a) Volume rendering of the vortex core with (b--d) contour plots of the total streamwise vorticity. Vortex cores are outlined in black.}
\label{fig:vorticity-contour}
\end{figure}

We study the decay of the vorticity shed from yawed wind turbines in the (neutrally-stratified) ABL using LES of yawed actuator disks. LES is  performed with the pseudo-spectral/finite difference code LESGO, which has been used and validated in previous work~\citep{Calaf2010a,Stevens2018a}. The coordinate system $\boldsymbol{x} = (x,y,z)$ with the unit vectors $\boldsymbol{i}$, $\boldsymbol{j}$, and $\boldsymbol{k}$ is defined such that $x$ is the streamwise direction, $y$ is the spanwise direction, and $z$ is the vertical direction. The origin is placed at the center of the disk with radius $R = D/2 = 50$ m. The effective domain size is $L_x = 3.75$ km, $L_y = 3$ km, and $L_z = 1$ km, and we use  $360 \times 288 \times 432$ grid points. Turbulent inflow is generated using a concurrent precursor domain~\citep{Stevens2018a} with a friction velocity of $u_* = 0.45$ m/s. A shifted periodic boundary condition~\citep{Munters2016a} with a $0.49L_z$ shift  is used to reduce streamwise streaks in the time-averaged velocity field. The wind turbine with hub height $z_h = 100$ m is placed 500 m downstream of the domain inlet. Subgrid stresses are modeled using the Lagrangian-averaged scale dependent model~\citep{Bou-Zeid2005a}. Wall stresses are modeled using the equilibrium wall model~\citep{Moeng1984a} with roughness length $z_0 = 0.1$ m.

The wind turbine is treated as a porous actuator disk that exerts an axial force 
$T = -\frac{1}{2} \rho \pi R^2 C_T' u_d^2$, perpendicular to the disk, that depends on the local thrust coefficient $C_T'$, disk-averaged velocity $u_d$, disk radius $R$, and the density of air $\rho$. The total axial force $T$ is distributed across the disk, leading to a distributed force $\boldsymbol{f}(\boldsymbol{x}) = T \, \mathcal{R}(\boldsymbol{x}) \boldsymbol{n}$, using the normalized indicator function $\mathcal{R}(\boldsymbol{x})$, and points in the unit normal direction to the disk $\boldsymbol{n}$. The yaw angle $\gamma$ is measured counter-clockwise from the positive $x$-axis toward the positive $y$-axis such that the unit normal of the actuator disk is $\boldsymbol{n} = \cos \gamma \, \boldsymbol{i} + \sin \gamma \, \boldsymbol{j}$.  The normalized indicator function $\mathcal{R}=G*\mathcal{I}$ is found by filtering (convolving) $\mathcal{I}(\boldsymbol{x}) = \pi^{-1} R^{-2}\delta(x) H(R-r)$ (where $\delta(x)$ is the Dirac delta function and $H(x)$ is the Heaviside function) with a  filtering function $G$. The latter is a three-dimensional Gaussian whose width $\sigma_\mathcal{R} = \Delta / \sqrt{12}$ is equivalent to a top-hat filter~\citep{Pope2000a} with a filter size chosen as $\Delta = 1.5h$, where $h = ( \Delta x^2 + \Delta y^2 + \Delta z^2)^\frac{1}{2}$ is the root mean square of the grid spacings.

Simulations are run for yaw angles of $\gamma = 15^\circ$, $20^\circ$, $25^\circ$, and $30^\circ$ with a local thrust coefficient of $C_T' = 1.33$. Velocity fields are time-averaged for a time ${\cal{T}}$ where ${\cal{T}} u_*/L_z \approx 8$ (all variables in this paper are  time-averaged). A representative time-averaged streamwise vorticity $\omega_x$ field for $\gamma = 20^\circ$  is shown in Figure~\ref{fig:vorticity-contour}. The vorticity contour plots and volume rendering show  the initial generation of arcs of vorticity above and below the turbine line of symmetry. These arcs decay downstream, each tending to a more axisymmetric distribution. The bottom vortex becomes flattened, presumably due to the action of the ground. Furthermore, secondary vortex structures are generated at the ground.

\begin{figure}
\centerline{\includegraphics[width=\textwidth]{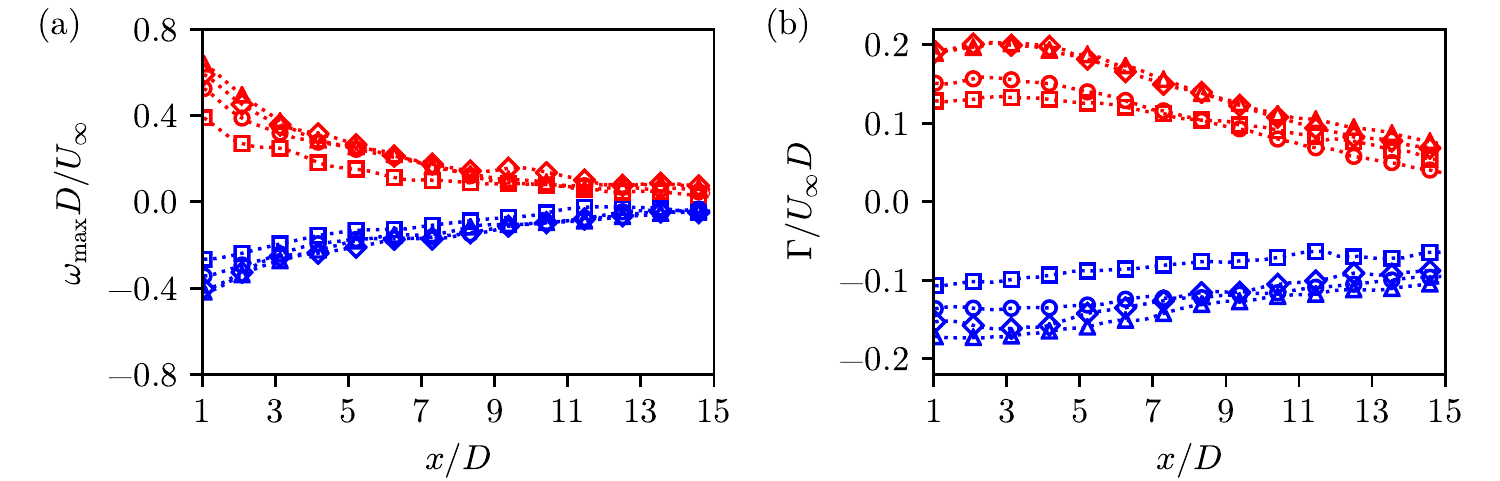}}
\caption{(a) Maximum vorticity magnitude and (b) circulation magnitude for top (blue and negative) and bottom (red and positive) vortices with $\gamma = 15^\circ$ ($\square$), $20^\circ$ ($\circ$), $25^\circ$ ($\diamond$), and $30^\circ$ ($\smalltriangleup$).}
\label{fig:decay-tb}
\end{figure}

Even with the significant time-averaging and shifted periodic boundary conditions of the inflow, some  background (noisy) vorticity is evident in the contour plots. To distinguish between the shed CVP and the background vorticity, we apply Otsu's method~\citep{Otsu1979a} on the positive and negative vorticity at each cross-plane. Otsu's method maximizes the intercategory (or minimizes the intracategory) variance, and thus identifies the region with the strongest coherent vorticity, which we define as the vortex core. 

To determine the circulation of each vortex as function of $x$, we numerically integrate the vorticity over the core area to obtain  $\Gamma_\mathrm{core}(x)$. The core vorticity ratio $\alpha(x) = \omega_\mathrm{Otsu}(x)/\omega_\mathrm{max}(x)$ is defined as the ratio of the thresholding value on vorticity that separates the core vortex region from the remaining vorticity $\omega_\mathrm{Otsu}(x)$ to the maximum vorticity magnitude $\omega_\mathrm{max}(x)$. The total circulation of each vortex is then estimated as $\Gamma(x) =\Gamma_\mathrm{core}(x)/(1-\alpha(x))$. This approach exactly recovers the total circulation of a Lamb-Oseen vortex~\citep{Saffman1992a}. The downstream evolution of maximum vorticity magnitude $\omega_\mathrm{max}(x)$ and circulations $\Gamma(x)$ measured from LES and normalized by the inlet velocity $U_\infty$ and disk diameter $D$ are shown in Figure~\ref{fig:decay-tb}. We see similar decaying behaviour for all yaw angles with the bottom vortex initially having a greater core circulation than the top vortex and decaying more quickly. Unlike the peak vorticity that begins to decay immediately downstream of the turbine, the circulation stays nearly constant up to $x/D\sim 3$ and only then begins its decay downstream. 

A number of vortex decay mechanisms have been discussed~\citep{Jaarsveld2011a}, such as viscous diffusion, strong external turbulence, cross-diffusion across the line of symmetry, and Crow instability breakup. When turbulence levels and shed vorticity strength are moderate, evidence from many of these earlier simulations points to the cross-diffusion mechanism~\citep{Cantwell1988a, Ohring1993a, Dommelen1995a} playing a dominant role. In the following sections, we develop a model first to predict the generation and then the decay of the yawed wind turbine CVP.

\section{Generation of counter-rotating vortices from yawed actuator disks}
\label{sec:generation}
We first model the generation of the vorticity at the rotor plane. By approximating the elliptic projection of the transverse force of an actuator disk as a circle, the transverse force can be written as 
\begin{equation}
f_y = -\textstyle{\frac{1}{2}}\rho \, C_T U_\infty^2 \,\cos^2 \gamma   \, \sin \gamma  \, H(R-r) \delta (x),
\label{eq:fy}
\end{equation}
where $C_T$ is the standard thrust coefficient and $r$ is the radial distance along the disk. We also use $r$ written in terms of the transverse coordinates (i.e. $r^2 = y^2 + z^2$), and $\theta$ is the polar angle measured from the positive $y$-axis toward the positive $z$-axis, i.e.,  $\sin \theta = z/r$. 
Taking the curl of the mean  momentum equation, linearizing the advective term, and neglecting turbulent and viscous stresses, the linearized mean streamwise vorticity transport equation (also used in~\cite{Martinez2017b}) becomes
\begin{equation}
\label{eq:curl_mom}
  U_\infty \partial_x \omega_x= - \rho^{-1} \partial_z f_y.
\end{equation}
Writing the derivative of the transverse force in terms of the cylindrical coordinate system using the chain rule, using~\eqref{eq:fy} and integrating~\eqref{eq:curl_mom} yields the vorticity distribution
\begin{equation}
\label{eq:vorticity}
\omega_x(x,r,\theta) = -\textstyle{\frac{1}{2}} C_T U_\infty \cos^2\gamma \,\sin \gamma \sin \theta \,\, \delta(r-R) H(x).
\end{equation} 
Integration of the vorticity~\eqref{eq:vorticity} just downstream of the disk over the top and bottom half-planes  yields the circulation of both top and bottom shed vortices
\begin{equation}
\Gamma_\mathrm{top} \! = \! -\Gamma_\mathrm{bottom}\! = \!\int_0^\infty \!\! \int_{0}^{\pi} \!\!\omega_x(0^+,r,\theta) \, r \, d\theta \, dr \! = \! -R C_T U_\infty \cos^2 \gamma \sin \gamma\! .
\end{equation}
The vortices are counter-rotating with a circulation magnitude $\Gamma_0 = R C_T U_\infty \cos^2 \gamma \sin \gamma$ identical to the predictions from lifting line theory~\citep{Shapiro2018b}.

\begin{figure}
\centerline{\includegraphics[width=0.84\textwidth]{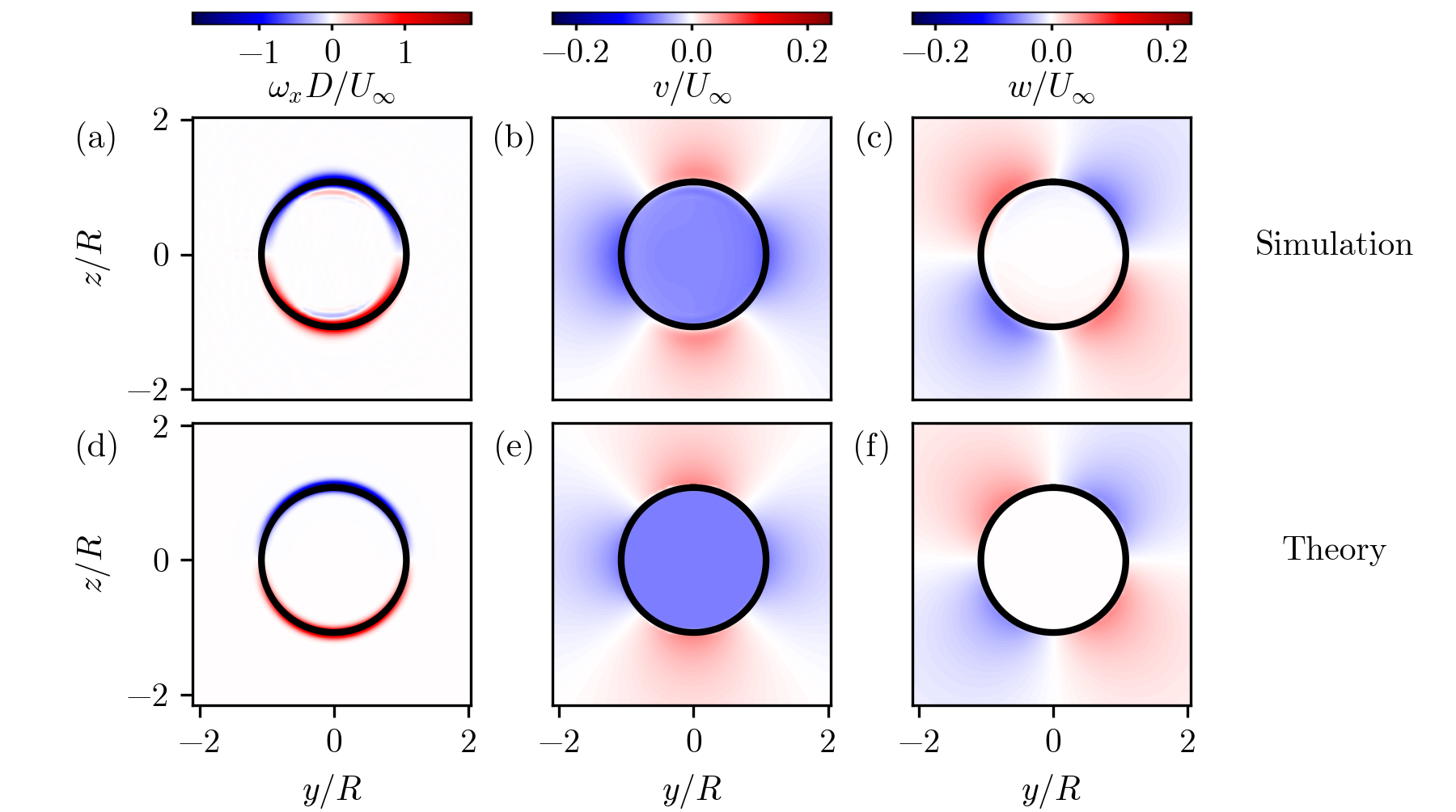}}
\caption{(a,d) Near rotor streamwise vorticity, (b,e) spanwise velocity, and (c,f) vertical velocity distributions from a yawed actuator disk with $C_T' = 0.8$ and $\gamma = 20^\circ$, with laminar inflow. Top panels show values measured at $x=R$ and bottom panels show theory. A circle with radius $R_*$ is shown in black in all panels.}
\label{fig:generation}
\end{figure}

The vorticity predicted by~\eqref{eq:vorticity}, which is valid for an idealized actuator disk, is now compared to numerical simulations of a yawed actuator disk under uniform inflow from~\cite{Shapiro2018b}. The vorticity distribution under the filtered forcing in these simulations can be approximated by first mapping~\eqref{eq:vorticity} with an effective radius  $R_* = R + 0.75h$~\citep{Shapiro2018b} and circulation $\Gamma_0^* = R_* C_T U_\infty \cos^2 \gamma \sin \gamma$ onto an arc shaped line, where $\omega_x(\chi,\zeta) =- \Gamma^*_0/(2R_*) \sin\left(\chi/R_*\right) \delta (\zeta)$, $\chi = \theta r$, and $\zeta = r-R_*$. This vorticity is then filtered (convolved) with a two-dimensional Gaussian $G_2 = (2\pi\sigma_\mathcal{R}^2)^{-1}\exp(-(\chi^2+\zeta^2)/2\sigma_R^2)$ whose width $\sigma_\mathcal{R}$ is equal to the filtering kernel used to filter the axial force to obtain
\begin{equation}
\label{eq:vorticity-line-sine}
\omega_x(\theta,r) = -\frac{\Gamma^*_0}{2R_*} \frac{\sin (\theta r/R_*)}{\sigma_\mathcal{R} \sqrt{2 \pi}} \mathrm{exp}\left( - \frac{(r-R_*)^2}{2 \sigma_\mathcal{R}^2} \right)  \mathrm{exp}\left( - \frac{\sigma_\mathcal{R}^2}{2R_*^2} \right).
\end{equation}
As can be seen in Figure~\ref{fig:generation}  for the case with $C_T' = 0.8$ and $\gamma = 20^\circ$, the vorticity distribution predicted by~\eqref{eq:vorticity-line-sine}, Figure~\ref{fig:generation}(a), reproduces the numerical results, Figure
~\ref{fig:generation}(b), with the simulation performed for the same parameters.
For comparison to simulations, the thrust coefficient is calculated based on the local one used for the simulations according to $C_T = 16 C_T'/(4+C_T'\cos^2\gamma)^2$~\citep{Shapiro2018b}.

To  validate the vorticity generation model, we also compare induced velocities by applying the Biot-Savart law in the near turbine region:
\begin{align}
v(\boldsymbol{x}) = -\frac{1}{4\pi} \int \frac{\omega_x(\boldsymbol{x}') (z - z')}{\vert \boldsymbol{x} - \boldsymbol{x}'\vert^3} d^3 \boldsymbol{x}'  \qquad 
w(\boldsymbol{x}) = \frac{1}{4\pi} \int \frac{\omega_x(\boldsymbol{x}') (y - y')}{\vert \boldsymbol{x} - \boldsymbol{x}'\vert^3} d^3 \boldsymbol{x}'.
\end{align}
Integrating in the radial direction we obtain
\begin{equation}
v(\boldsymbol{x}) \! =  \!\frac{1}{8\pi}  \frac{\Gamma_0}{R}  \int_{0}^{\infty} \! \! \!\int_{0}^{2 \pi} \! \! \frac{ R \sin \theta'  (r \sin \theta - R\sin \theta')  \, d\theta'  \, dx'}{\left[(x-x')^2 + (r\cos \theta - R \cos \theta')^2  + (r\sin \theta - R \sin \theta')^2\right]^{\frac{3}{2}}},
\end{equation}
and integration in the streamwise direction~\citep[\#2.271.5]{Gradshteyn1980a} yields 
\begin{equation}
v(\boldsymbol{x}) = \frac{1}{8\pi}  \frac{\Gamma_0}{R}  \int_{0}^{2 \pi}  R \sin \theta'  (r \sin \theta - R\sin \theta') \left[ \frac{1}{a} +   \frac{1}{a} \frac{x}{(a + x^2)^{1/2}} \right]\, d\theta',
\end{equation}
where  $a = (r\cos \theta - R \cos \theta')^2  + (r\sin \theta - R \sin \theta')^2$. We are primarily interested in   $v$ at $x >>R$ or $x>>a$, leading to $v=-\Gamma_0/4R$ and $w=0$ for $r \le R$, and $v = -{\Gamma_0}/({4R}) (R/r)^2\cos(2\theta)$ and $w = -{\Gamma_0}/({4R}) (R/r)^2\sin(2\theta)$ for $r > R$. The $w$ component has been found by using the continuity equation. 
Inside the radius of the actuator disk, the $v$ velocity component ($\Gamma_0/4R$) is identical to the constant prediction from lifting line theory \citep{Shapiro2018b}, and the $w$ component vanishes.  Outside the radius of the actuator disk, the velocity components depend on the polar angle and decrease with the squared radial distance. 

The  predictions for $v$ and $w$ 
are compared to simulations for $C_T' = 0.8$ and $\gamma = 20^\circ$ measured at $x = R$ in Figure~\ref{fig:generation}(b-c,e-f). To compare the theoretically predicted velocity components to simulation results, the velocity must be sampled before the self-induction of the vorticity is considerable. However, directly downstream of the actuator disk, the actuator disk streamtube is still expanding from the non-negligible streamwise pressure gradient induced by the streamwise component of the axial force. To counteract this effect in the simulation measurements, we have removed the expansion expected from a decelerating streamtube by plotting $v + u_r\cos\theta$ and $w + u_r \sin \theta$, where $u_r$ is the radial velocity. It is obtained by measuring the streamwise velocity gradient at the center of the actuator disk streamtube (assuming that $\partial_x  u= \partial_x u(R,0,0)$ for $r \le R^*$ and $\partial_x u = 0$ for  $r > R^*$) and radially integrating the continuity equation, i.e. $u_r = ({r}/{2}) \partial_x u(R,0,0)$ for $r \le R^*$ and  $u_r = ({R_*^2}/{2r}) \partial_x u(R,0,0)$ for $r > R_*$. With this correction included, the velocity components agree well with simulations, thus further supporting the predicted generated vorticity distribution as in~\eqref{eq:vorticity}.

\section{Turbulent decay of counter-rotating vortices in the ABL}
\label{sec:decay}
We  now consider the decay of the CVP due to the surrounding turbulence in the ABL and test the implications of the cross-diffusion hypothesis~\citep{Cantwell1988a, Ohring1993a, Dommelen1995a}. In our simplified model, the self-induced deformation of the shed vorticity sheet is neglected, the ABL shear is also neglected, and only turbulent diffusion is considered. The boundary layer assumptions are applied to the streamwise vorticity equation downstream of the turbine~\citep{Saffman1992a, Pope2000a}, and~\eqref{eq:curl_mom} is replaced by an advection-diffusion equation with eddy viscosity $\nu_T(x)$:
\begin{equation}
\label{eq:adv-diff}
U_\infty \partial_x \omega_x = \nu_T(x)  \left(\partial_y^2 \omega_x +  \partial_z^2 \omega_x \right).
\end{equation}

First, note that a point vortex $\omega_x(x_0,y,z) = \Gamma_p \delta(y-y_0) \delta(z-z_0)$ with circulation $\Gamma_p$ located at $(x_0,y_0,z_0)$ that evolves under~\eqref{eq:adv-diff} diffuses downstream~\citep{Saffman1992a} as
\begin{equation}
\label{eq:point-vorticity-diffused}
\omega_x(x,y,z) = \frac{\Gamma_p}{4 \pi \eta^2(x)}\exp\left(-\frac{(y-y_0)^2 + (z-z_0)^2}{4 \eta^2(x)}\right),
\end{equation}
where the viscous length scale $\eta(x)$ results from the integral of the eddy viscosity
\begin{equation}
\eta^2(x) = U_\infty^{-1} \begingroup\textstyle \int\endgroup_{x_0}^x \nu_T(x') \, dx'.
\end{equation}
The virtual origin $x_0$ is introduced to account for the finite thickness of the initial vorticity distribution, which depends on the  grid size in simulations or potentially the chord size of a physical turbine.
The solution in  \eqref{eq:point-vorticity-diffused} is equivalent to filtering the initial condition with a two-dimensional Gaussian kernel with a width of $\sqrt{2}\, \eta(x)$. This result is then applied to the initial vorticity distribution~\eqref{eq:vorticity} generated by the yawed turbine by placing point vorticies around the circle with radius $R$ at locations $(x_0,R\cos\theta, R\sin\theta)$ with differential circulation $d\Gamma_p = -\Gamma_0\sin\theta/2 \, d\theta$. 
Integrating around the circle leads to total vorticity
\begin{equation}
\label{eq:vorticity-diffused}
\omega_x(x,y,z) = -\int_0^{2\pi} \frac{\Gamma_0 \sin\theta}{8 \pi \eta^2(x)}\exp\left(-\frac{(y-R\cos\theta)^2 + (z-R\sin\theta)^2}{4 \eta^2(x)}\right) \, d\theta.
\end{equation} While~\eqref{eq:vorticity-diffused} cannot be integrated directly for all $y$ and $z$, the integral of \eqref{eq:vorticity-diffused} coincident with the peak vorticity magnitude at $y=0$ and $z = \pm R$ can be integrated as
\begin{align}
\label{eq:max-vorticity}
\omega_\mathrm{max}(x) = \frac{\Gamma_0}{R^2} \frac{R^2}{4 \eta^2(x)} \exp\left( - \frac{R^2}{2\eta^2(x)}\right) I_1\left( \frac{R^2}{2\eta^2(x)}\right),
\end{align}
where $I_n$ is the modified Bessel function of the first kind with order $n$.

The total circulation in the vortex system generated by a yawed actuator disk vanishes in all streamwise planes, i.e. $\Gamma_\mathrm{total}(x) = \int_{-\infty}^\infty \int_{-\infty} ^ \infty \omega_x(x,y,z) \, dy \, dz = 0$, because the vorticity across the $y$-axis is equal and opposite. Integrating each vortex $\Gamma (x) =  \vert \int_{0}^\infty \int_{-\infty} ^ \infty \omega_x(x,y,z)  \, dy \, dz \vert =  \vert \int_{-\infty}^0 \int_{-\infty} ^ \infty \omega_x(x,y,z)  \, dy \, dz \vert$
yields a normalized circulation
\begin{equation}
\label{eq:circulation}
\frac{ \Gamma(x) }{\Gamma_0}= \frac{\sqrt{\pi}}{4}\frac{R}{\eta(x)} \exp\left( - \frac{R^2}{8\eta^2(x)}\right)\left[ I_0\left( \frac{R^2}{8\eta^2(x)}\right) + I_1\left( \frac{R^2}{8\eta^2(x)}\right)\right],
\end{equation}
whose magnitude monotonically decreases for $\eta \ge 0$. This decrease in circulation is caused purely by the cancellation of vorticity along the $y$-axis as vorticity diffuses downstream.

The problem of properly specifying the eddy viscosity is approached using a mixing length model $\nu_T(x) = \upsilon \ell$, where $\upsilon$ is a velocity scale and $\ell$ is the mixing length. In the ABL, the appropriate velocity scale is the friction velocity, i.e. $\upsilon = u_*$. From similarity scaling for a wake in the boundary layer~\citep{Shapiro2019b}, we know that a wake will grow linearly with downstream distance, i.e. $\ell \sim x$. Equivalently, the Jensen wake model~\citep{Jensen1983a} assumes that the diameter of a top-hat wake is $D_w = D + 2kx$, where $k$ is the wake expansion rate commonly taken as $k = u_*/U_\infty$. We assume that the vorticity grows at the same rate $2kx$, but initially starts with thickness much smaller than $D$. In order to write $\ell$ in terms of the Jensen model top-hat length scale, we note that a point vortex filtered with a box filter with a scale $\beta$ has the same second moment~\citep{Pope2000a} as a viscously diffused point vortex with length scale $\beta/\sqrt{24}$. Therefore, we write the mixing length as $\ell = 2 k x/\sqrt{24}$. Thus the resulting eddy viscosity and squared viscous length scale are respectively modeled according to
\begin{equation}
\label{eq:viscous-length}
\nu_T(x) = u_* 2k(x-x_0)/\sqrt{24}  \qquad\mbox{and}\qquad  \eta^2(x) = k^2 (x-x_0)^2 /\sqrt{24}.
\end{equation}

\begin{figure}
\centerline{\includegraphics[width=\textwidth]{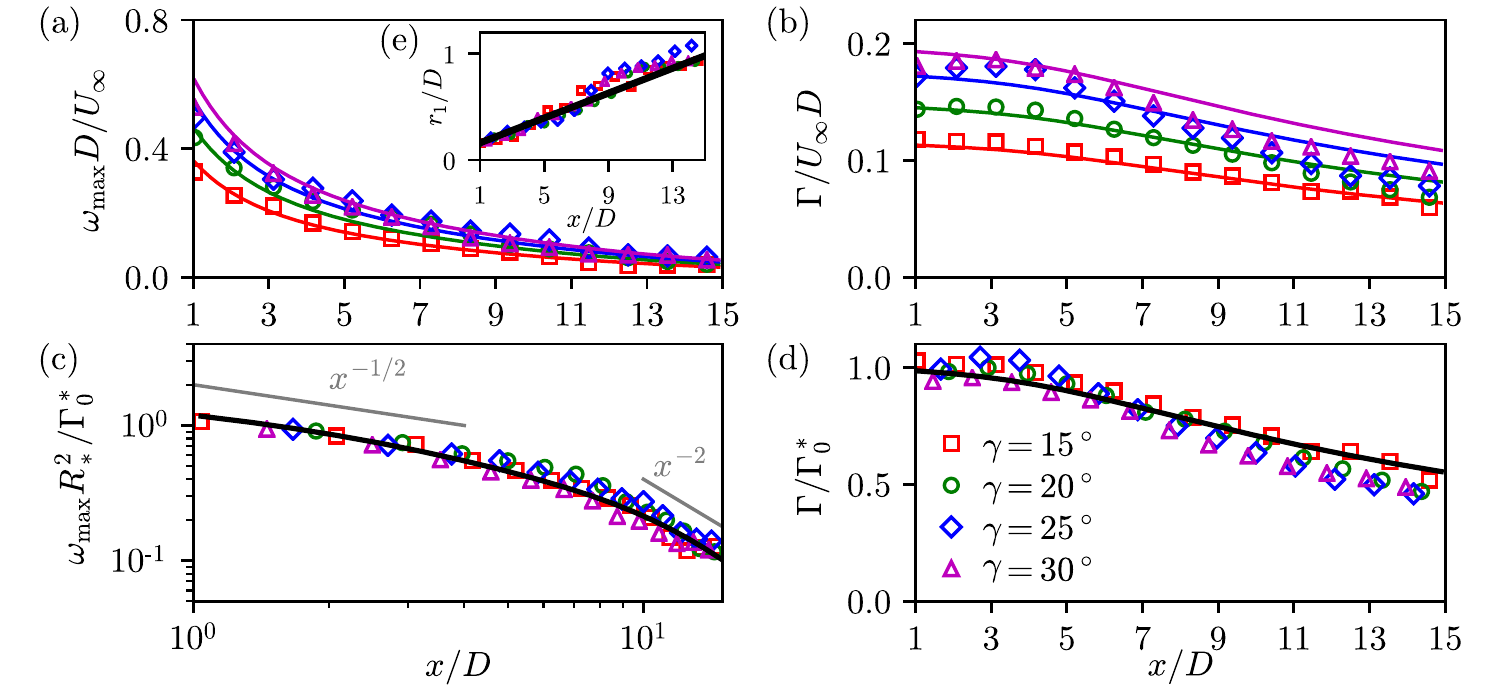}}
\caption{Maximum vorticity and circulation magnitude normalized by (a--b) rotor diameter $D$ and inlet velocity $U_\infty$ and (c--d) theoretical circulation $\Gamma_0^*$ and effective radius $R_*$. (e) Vortex radius showing linear growth. Symbols show simulation results (the arithmetic average of the magnitudes corresponding to top and bottom vortices) and lines show theory, i.e.~\eqref{eq:max-vorticity} for (a,c),~\eqref{eq:circulation} for (b,d), and~\eqref{eq:rone} for (e).}
\label{fig:decay}
\end{figure}

The maximum vorticity, circulation, and vortex growth rate are now compared to data from simulations in Figure~\ref{fig:decay}. In the model, the virtual origin $x_0$ is chosen by noting the equivalence between the effect of viscous diffusion with a length scale $\eta$  to Gaussian filtering with a length scale $\sqrt{2}\,\eta$. Considering the filtered axial force with length scale $\sigma_\mathcal{R} = \Delta/\sqrt{12}$, we conclude that the virtual origin is $x_0 = 24^{-1/4}\Delta/k$. We compare the model predictions with the arithmetic average of LES measured peak vorticity and circulation magnitudes from the top and bottom vortices, since the simulation data showed some differences between the top and bottom vorticies and these differences are not included by the current theory. Results shown in Figure~\ref{fig:decay} (c,d) also show that the normalizations by $R_*$ and $\Gamma_0^*$ suggested by the theory for maximum streamwise  vorticity~\eqref{eq:max-vorticity} and circulation~\eqref{eq:circulation} (with effective parameters in LES $R_*$ and $\Gamma_0^*$ determined as explained in \S \ref{sec:generation}) yield good collapse of the LES data and with the theory. 

Consideration of the transformation of the vorticity from a diffused line around the edge of the disk to a diffused point vortex as well as the viscous length scale $\eta(x)$ reveals power law scalings for the maximum vorticity. With the virtual origin from the simulations $x_0/D \approx 2$, the viscous length scale $\eta(x)$ initially scales as $\eta(x) \sim x$ for $x/D<4$. In the far field, $x/D>10$, the scaling changes to $\eta(x) \sim x^2$. Initially, the vorticity is confined to a line around the edge of the disk, giving an expected scaling of $\omega_\mathrm{max}(x) \sim \eta^{-1/2} \sim x^{-1/2}$. Conversely, in the far field, the vorticity behaves like a point vortex with the expected scaling of $\omega_\mathrm{max}(x) \sim \eta^{-1} \sim x^{-2}$. These scaling laws agree well with simulations and theory, as shown in Figure~\ref{fig:decay}(c). Finally, to validate the growth rate of the mixing length, we calculate the vortex radius from simulations, which is defined as the location of the maximum spanwise velocity above the rotor $r_1(x) +R = \mathrm{argmax} \, v(x,0,z)$. For a Lamb-Osseen vortex, which is expected beyond $x/D > 5$, the vortex radius is~\citep{Saffman1992a} 
\begin{equation}
r_1 = 2.24 \eta = 2.24 \, (24)^{-\frac{1}{4}} k(x-x_0) \approx k(x-x_0),  ~~~{\rm with} ~~~k=u_*/U_\infty.
\label{eq:rone}
\end{equation}
As shown in Figure~\ref{fig:decay}(e), the growth rate of the measured vortex radius is linear with $x$ and agrees with the theory. 

\section{Discussion and Conclusions}
Using concepts drawn from the airplane trailing vortex literature~\citep{Cantwell1988a, Ohring1993a, Dommelen1995a, Spalart1998a}, we study  the decay of the vortices generated by yawing of wind turbines. The theory presented in \S\ref{sec:generation} and \S\ref{sec:decay} considers the effect of linear advection and turbulent diffusion on the  decay of the vorticity and circulation shed from yawed turbines. The analysis is based on a streamwise-varying eddy viscosity that depends on the growth rate of the vorticity length scale and the boundary layer friction velocity. 
The analysis enables us to obtain analytical expressions for the maximum vorticity and shed circulation from each of the CVP that agree well with actuator disk simulations of yawed wind turbines in the ABL. Results refine the emerging understanding of the decay of the vorticity shed from yawed turbines. As in~\cite{Shapiro2019b}, we find that careful consideration of the appropriate mixing length and velocity scale for the eddy viscosity of wind turbine in the ABL, yields an eddy viscosity that increases linearly with downstream distance and a mixing length that grows at a rate $k=u_*/U_\infty$. Also, the results provide a theoretical framework for engineering models of the shed vorticity consisting of closed-form analytical expressions, 
i.e. \eqref{eq:max-vorticity}, \eqref{eq:circulation} and~\eqref{eq:rone}. These do not require numerical integration of differential equations to evaluate the model, hence facilitating eventual use in engineering models for wind farm design and control. The scaling also agrees well with the empirical observation of~\cite{Zong2020a} in the near field of the wake. 

Turbulent mixing appears to be the dominant process that governs the decay of the shed vorticity. The yawed turbine generates equal and opposite circulation bound to the rotor disk that is shed downstream, resulting in vanishing total circulation. For a single vortex the circulation would remain constant  even as the vorticity diffuses downstream. However, since the opposing negative vorticity similarly diffuses, the cancellation of the diffused vorticity along the centerline of the wake results in the apparent ``dissipation" of circulation for the entire system. The cross-diffusion hypothesis, however, does not fully explain the apparent differences between the top and bottom vortices in the CVP. Ground effects, vertical shear, and the vertical structure of turbulence in the ABL clearly play a role in creating some differences in the evolution of the top and bottom vortices that more refined models should also aim to reproduce.

\section*{Acknowledgements}
The authors acknowledge funding from the National Science Foundation (grant nos. 1949778 and 1635430) and computational resources from MARCC and Cheyenne (doi:10.5065/D6RX99HX).
 
\section*{Declaration of interests}
The  authors  report  no  conflict  of  interest.

\bibliographystyle{jfm}
\bibliography{main}

\end{document}